\documentclass[a4paper]{jpconf}
\usepackage{graphicx}
\usepackage{verbatim}
\usepackage{amsmath}
\usepackage{iopams}
\usepackage[final]{changes}
\setremarkmarkup{\footnote{#2}}
\setsocextension{changes}
\definechangesauthor[name={referee replies}, color=red]{RW}
\definechangesauthor[name={later added comments}, color=blue]{PS}

\begin{document}
\twocolumn[
\title{Relaxation to magnetohydrodynamics equilibria via collision brackets}

\author{C~Bressan$^{1,2}$, M~Kraus$^{1,2}$, P~J~Morrison$^{3}$ and O~Maj$^{1,2}$}  %
\address{$^1$ Max-Planck-Institute for Plasma Physics, Garching, Germany} 
\address{$^2$ Technische Universit\"at M\"unchen, Zentrum Mathematik, Garching, Germany} 
\address{$^3$ The University of Texas at Austin, Physics Department and Institute for Fusion Studies, USA} 

\ead{camilla.bressan@ipp.mpg.de}

\begin{abstract}
Metriplectic dynamics is applied to compute equilibria of fluid dynamical
systems. The result is a relaxation method in which Hamiltonian dynamics
(symplectic structure) is combined with dissipative mechanisms (metric
structure) that relaxes the system to the desired equilibrium point. The specific metric operator, which is considered in this work, is formally analogous to the Landau collision operator. These ideas are illustrated by means of case studies. 
The considered physical models are the Euler equations in vorticity form, the Grad-Shafranov equation, and force-free MHD equilibria.
\end{abstract} 
]

\section{Introduction}\label{sec:introduction} 
The computation of general 3D magnetohydrodynamic (MHD) equilibria plays a fundamental role in simulations of stellarators and it is important for tokamaks as well, due to deviations from axisymmetry (magnetic islands, ripples, and resonant magnetic perturbations). \\
This problem has always attracted interest in the plasma physics community, leading to different numerical approaches \cite{vmec1, vmec2, pies, siesta, spec}. Nonetheless, the efficient computation of three-dimensional MHD equilibria is still an open issue.

In the geometric mechanics community, on the other hand, there has been a significant effort directed to the study of the appropriate geometric structures for the description of dissipative systems and irreversible dynamics. Such a structure has been proposed by Morrison \cite{Morrison1984, Morrison1986} and it is referred to as metriplectic dynamics since it combines the symplectic structure of Hamilton's equations with the metric structure of gradient flows. Several physical systems can be cast in metriplectic form, e.g., the free rigid body with suitable chosen torque \cite{MaterassiMorrison}, resistive MHD \cite{MaterassiAndTassi} and the Lindblad equation for open quantum systems \cite{MittnenzweigAndMielke}. However, the geometric properties of metriplectic flows can also be exploited to design artificial dynamical systems that relax to an equilibrium of the considered physical system.
The advantages of such methods come from properties directly implied by the geometric structure and the energy-Casimir principle \cite{PhilReviewPaper, Gay-BalmazHolm, Guy-Balmaz_selectivCasimir_MHD}.  

A related but different approach has been proposed by Flierl and Morrison \cite{FlierlAndMorrison} and developed further by Chikasue, Furukawa and Morrison \cite{ChikasueFurukawa1, ChikasueFurukawa2, FurukawaMorrison3, FurukawaMorrison4}. In such an approach, the relaxation mechanism is constructed on the basis of the symplectic structure only, essentially by squaring the Poisson operator. This method has the properties of minimizing the energy functional of the system, while preserving all the other Casimir invariants; for the case of ideal MHD this implies the preservation of the \added[id=PS, remark=clearer statement]{magnetic field line} topology determined by the initial conditions. %

The present work is set in the framework of metriplectic dynamics. A specific metric operator is constructed on the lines of the Landau operator for Coulomb collisions, which has a metriplectic structure already discovered by Morrison \cite{Morrison1984, Morrison1986}. %
The basic idea is developed for three case studies in order to explore the advantages and disadvantages of the proposed relaxation method.

\section{Theory}\label{sec:theory} 
For a class of dynamical systems arising in fluid and kinetic theories of plasmas, equilibrium states can be characterised by a variational principle.

Typically equilibrium states are extrema of an entropy functional under the constraints imposed by the first integrals of the systems, such as mass, energy or momentum. For instance, in the Boltzmann equation (with collisions) equilibria are obtained by extremizing the entropy at constant energy, momentum, and particle number \cite{Lenard}. Moreover, the Boltzmann equation has the additional property that a solution of the initial value problem relaxes, as time goes to infinity, to an equilibrium because of the celebrated H theorem \cite{Lenard}; equilibria can therefore be identified by time-evolution of properly chosen initial conditions. In general this is not the case: ideal systems with no dissipation mechanisms will not relax to an equilibrium. Therefore, in order to design a relaxation method for the computation of equilibria, some dissipation mechanism has to be introduced.

The idea proposed by Morrison \cite{Morrison1984} shows the possibility to define a dissipative dynamics that relaxes to a solution of the variational problem for the equilibrium. These concepts will be explained in more detail here with the help of specific physical models.

\subsection{Physical Models}\label{subsec:physical_models} 
Three specific case studies are presented in order to illustrate the proposed idea.

The first example uses the vorticity form of the 2D Euler equations, 
\begin{equation}\label{eq:euler_model} 
  \begin{split}
    \partial_t \omega(t, x) + [\omega(t, x), \phi(t, x)] &= 0 \\
    -\Delta \phi(t, x) &= \omega(t, x),
  \end{split} 
\end{equation}
where $t$ is time and $x=(x_1, x_2)$ are the Cartesian coordinates in the two-dimensional space. The dynamical variable $\omega$ is the vorticity of an incompressible flow $v = (\partial_y \phi, -\partial_x \phi)$ in two-dimensions and $\phi$ is the stream function. The two-dimensional Laplacian is $\Delta = \partial^2_{x_1} + \partial^2_{x_2}$ and $[f, g] = \partial_{x_1} f \partial_{x_2} g - \partial_{x_1} g \partial_{x_2} f$, for any pair of functions $f,g$. %
The equilibrium states of the Euler system are reached when $\partial_t \omega = 0$. Then from $[\omega, \phi] = 0$, the vorticity must be proportional to a function of the stream function, $\omega = \lambda f(\phi)$. Substituting this expression into the Poisson equation of ~\eqref{eq:euler_model} leads to the non-linear eigenvalue problem
\begin{equation}\label{eq:euler_equilibrium}
-\Delta \phi = \lambda f(\phi),
\end{equation}
for the pair $(\lambda, \phi)$; given a solution, the corresponding vorticity field is determined by $\omega = \lambda f(\phi)$.
Complemented with boundary conditions, the nonlinear eigenvalue problem ~\eqref{eq:euler_equilibrium} determines a whole class of equilibrium states $(\lambda, \phi)$: for every choice of $f$, each solution $(\lambda, \phi)$ corresponds to an equilibrium.
\replaced[id=PS,remark=better formulation]{In the second case study, we apply}{The second uses} the method to solve the Grad-Shafranov equation \cite{GS1, GS2}:  
\begin{equation} \label{eq:gs_model} \begin{split}
-\Delta^* \psi (R, z) &=  \lambda f(\psi, R, z),  %
\end{split} \end{equation}
where $\Delta^* = R \partial_R \big(R^{-1} \partial_R  \big) + \partial^2_z,$ and
$(R, z)$ are the radial and axial coordinates of a cylindrical reference system
$(R,z,\varphi)$. Physically the unknown $\psi$ is a flux function and the right-hand side is $\lambda f(\psi,
R, z) = (4\pi/c) R j_{\varphi}$ with $j_{\varphi}$ the $\varphi$-component of the current density, and $c$ the speed of light in vacuum (c.g.s. units).

The Grad-Shafranov equation is formally analogous to equation~\eqref{eq:euler_equilibrium}, if the Laplace operator and the stream function $\phi$ are replaced by $\Delta^*$ and the flux function $\psi$, respectively. The same considerations about the equilibrium states apply.

As a last example, force-free MHD equilibria (also known as Beltrami fields or Taylor-relaxed states) are considered. The magnetic field $B = \big( B_1(x), B_2(x), B_3(x) \big)$, where $x = (x_1, x_2, x_3)$ are the Cartesian coordinates in three-dimensional domain, satisfies the force-free equilibrium condition if \added[id=PS,remark=better formulation]{it satisfies the Beltrami equation}
\begin{equation}\label{eq:beltrami_equilibrium}
  \nabla \times B = \mu B, \quad  B \cdot \nabla \mu = 0,
\end{equation}
where $\mu$ is in general a function.
If $\mu$ is constant, the Beltrami equation \eqref{eq:beltrami_equilibrium} reduces to the eigenvalue problem for the curl operator.

\subsection{Variational Principle}\label{subsection:variational_principle}
The equilibria of the considered systems can be characterised as the extrema of an entropy functional with the constraint \replaced[id=PS,remark=better formulation]{that a given Hamiltonian functional is preserved}{of constant Hamiltonian}.

We consider either the case of a scalar field $u = u(x)$ or a multicomponent
field $u  = \big(u_1(x), \ldots, u_n(x) \big)$, defined over a spatial domain $\Omega$. Let $\mathcal{S} = \mathcal{S}(u)$ and $\mathcal{H} = \mathcal{H}(u)$ be the entropy and Hamiltonian functionals, respectively. The problem of finding the extrema of $\mathcal{S}$ at constant $\mathcal{H}$ is written as \added[id=PS,remark={the sign of the Lagrange multiplier $\lambda$ has been changed for consistency and to underline the proportionality relationship between the functional derivatives of $\mathcal{S}$ and $\mathcal{H}$}]{}

\begin{equation}\label{eq:energy_casimir_principle}
\replaced{ 
\protect \frac{\delta \mathcal{S}(u)}{\delta u} = \lambda \frac{\delta \mathcal{H}(u)}{\delta u},
}
{
\protect \frac{\delta \mathcal{S}(u)}{\delta u} + \lambda \frac{\delta \mathcal{H}(u)}{\delta u} = 0,
 }
\end{equation} 

where $\lambda$ is the Lagrange multiplier.

In the Euler example, the dynamical variable $u$ is the scalar vorticity $\omega$. The energy functional $\mathcal{H}$ is the kinetic energy of the fluid (per unit mass) written as 
\begin{equation}\label{eq:euler_energy_functional} 
\mathcal{H}(\omega) = \frac{1}{2} \int_{\Omega} \omega(x)  \phi(x)  dx.
\end{equation}
We restrict the entropy functional to be of the form
\begin{equation}\label{eq:euler_entropy_functional} 
  \mathcal{S}(\omega) = \int_{\Omega} s(\omega(x)) dx,
\end{equation}
where $s = s(\omega)$ is smooth with monotonic derivative $s^\prime$. The functional derivatives are readily computed,
\begin{equation*}
  \frac{\delta \mathcal{H}(\omega)}{\delta \omega} = \phi, \quad
  \frac{\delta \mathcal{S}(\omega)}{\delta \omega} = s^\prime(\omega).
\end{equation*}
and equation ~\eqref{eq:energy_casimir_principle} becomes \added[id=PS,remark=consistent change of the sign of $\lambda$]{}
\begin{equation}\label{eq:energy_casimir_principle_euler}
\replaced{
s^\prime(\omega) = \lambda \phi.  %
}
{ s^\prime(\omega) + \lambda \phi = 0.
}
\end{equation}
One can now compare this result with equation ~\eqref{eq:euler_equilibrium}, and deduce a relationship between the choice of the entropy functional and a particular physical equilibrium described by the function $f$, namely,
\begin{equation}\label{eq:relation_physical_equilibrium_entropy}
f = (s^\prime)^{-1}.
\end{equation}
The inverse exists since $s^\prime$ is monotonic. This also implies that only equilibria with a monotonic $f$ can be described in this way.

In the case of the Grad-Shafranov equation, the natural variational principle
\cite{GS_standard_VP} seeks the extrema of an action functional written in terns of a Lagrangian density.
However, this variational principle is not in the form~(\ref{eq:energy_casimir_principle}).
In order to obtain a variational principle in the form~(\ref{eq:energy_casimir_principle}), let us introduce the variable $u = (4\pi/c)Rj_{\varphi}$. The flux function $\psi$ is determined from $u$ by solving the linear elliptic problem
\begin{equation*}
  -\Delta^* \psi = u,
\end{equation*}
equipped with the desired boundary conditions. %
Then we define the energy functional
\begin{equation}\label{eq:gs_energy_functional} 
  \mathcal{H}(u) = \frac{1}{2} \int_{\Omega} u(R, z)  \psi(R, z)  \frac{dR dz}{R},
\end{equation}
and we consider entropy functionals of the form
\begin{equation}\label{eq:gs_entropy_functional} 
  \mathcal{S}(u) = \int_{\Omega} s \big(u(R,z), R, z \big)  \frac{dR dz}{R}.
\end{equation}
As an example, the entropy
\begin{equation}\label{eq:gs-hm_entropy_functional}
  \mathcal{S}(u) = \frac{1}{2} \int_{\Omega}  \frac{u^2(R,z)}{CR^2 + D}  \frac{dR dz}{R},
\end{equation}
where $C$ and $D$ are positive constants, leads to the Herrnegger-Maschke solutions of the Grad-Shafranov equation \cite[and references therein]{GSMcCarthypaper}. Other choices, which will lead to different physical equilibria, can be made. The functional derivatives \added[id=PS,remark=relevant clarification: otherwise the functional derivatives would be different]{with respect to $L^2$ scalar product with metric $dRdz/R$} are
\begin{equation*}
  \frac{\delta \mathcal{H}(u)}{\delta u} = \psi, \quad
  \frac{\delta \mathcal{S}(u)}{\delta u} = s^\prime_u(u,R,z),
\end{equation*}
where $s^\prime_u(u) = \partial s(u,R,z)/\partial u$. Equation~(\ref{eq:energy_casimir_principle}) becomes \added[id=PS,remark=consistent change of the sign of $\lambda$]{}
\begin{equation*}
\replaced{
  s^\prime (u, R, z) = \lambda \psi.
}
{ s^\prime (u, R, z) + \lambda \psi = 0.
}
\end{equation*}
On using the entropy~\eqref{eq:gs_entropy_functional} for the sake of illustration, one obtains \added[id=PS,remark=consistent change of the sign of $\lambda$]{}  
\begin{equation}\label{eq:energy_casimir_principle_gs}
\replaced{
  \frac{u}{CR^2 + D} = \lambda \psi,
  }
  { \frac{u}{CR^2 + D} + \lambda \psi = 0,
  }
\end{equation}
and since $u = -\Delta^* \psi$, one obtains the weighted linear eigenvalue problem
\begin{equation*}
  -\Delta^* \psi = \lambda (C R^2 +D) \psi,
\end{equation*}
that characterises the Herrnegger-Maschke solutions.

As a last example, let us address Beltrami fields, that are the minimisers of the magnetic energy at constant magnetic helicity \cite{beltrami1, beltrami2}. 
Thus the \replaced[id=PS,remark=better formulation]{natural choice for the Hamiltonian functional is}{Thus the Hamiltonian functional should be} the magnetic helicity,
\begin{equation}\label{eq:beltrami_energy_functional} 
\mathcal{H} = \int_{\Omega} B(x) \cdot A(x) dx,
\end{equation}
where $A(x)$ is the magnetic vector potential. We fix the Coulomb gauge, 
\begin{equation}\label{eq:eqA}
  \nabla \times A = B,  \qquad
  \nabla \cdot A = 0. 
\end{equation}
With suitable boundary conditions, equation \eqref{eq:eqA} establishes a one-to-one relationship between $A$ and $B$. Correspondingly, the dissipated entropy is actually the physical energy of the magnetic field
\begin{equation}\label{eq:beltrami_entropy_functional} 
\mathcal{S} = \int_{\Omega}  \frac{|B|^2}{8\pi} dx.
\end{equation}
We have now two equivalent choices. The standard choice \cite{beltrami1} consists in setting $u=A$ and computing
\begin{equation*}
  \frac{\delta \mathcal{H}(A)}{\delta A} = 2 B, \quad
  \frac{\delta \mathcal{S}(A)}{\delta A} = \nabla \times B / (4\pi), 
\end{equation*}
and condition (\ref{eq:energy_casimir_principle}) gives the Beltrami equation directly with a constant
\replaced[id=PS,remark=consistent change of the sign of $\lambda$]{$\mu = +8\pi \lambda$}{$\mu = -8\pi \lambda$}.
Alternatively, one can set $u=B$, so that
\begin{equation*}
  \frac{\delta \mathcal{H}(B)}{\delta B} = 2 A, \quad
  \frac{\delta \mathcal{S}(B)}{\delta B} = B / (4\pi), 
\end{equation*}
and condition (\ref{eq:energy_casimir_principle}) reduces to \added[id=PS,remark=consistent change of the sign of $\lambda$]{}
\begin{equation*}
\replaced{
  B = 8\pi \lambda A, 
}
{ B + 8\pi \lambda A = 0,
}
\end{equation*}
which should be solved together with (\ref{eq:eqA}). Since $\lambda$ is a constant, this formulation is equivalent to the Beltrami equation. The latter choice appears more convenient in terms of computational cost.

\subsection{Metriplectic dynamics}
The metriplectic formulation of the dynamics of a (possibly multi-component) time-dependent field $u = u(t,x)$ reads: for every functional $\mathcal{F}$, the function $t \mapsto \mathcal{F}(u(t,\cdot) \big)$ must satisfy 
\begin{equation}\label{eq:joined_hamiltonian_dissipative_dynamics}
  \frac{d \mathcal{F}(u)}{dt} =  \{ \mathcal{F}(u), \mathcal{H}(u) \} + (\mathcal{F}(u), \mathcal{S}(u)),
\end{equation}
where $\{\cdot, \cdot\}$ is a Poisson bracket, that is an anti-symmetric, bilinear operation on the functionals, satisfying the Leibniz and Jacobi identities \cite{leibniz_jacobi}, whereas $( \cdot, \cdot)$ is a metric bracket, that is a symmetric, bilinear operation with a definite sign. In the following we shall assume that the metric brackets are negative semi-definite, but this is just a convention. The functional $\mathcal{F}$ plays the same role as the test-function in a weak formulation.

The Hamiltonian and entropy functionals must satisfy the conditions
\begin{equation*}
  \{ \mathcal{F}(u), \mathcal{S}(u) \} = 0, \qquad (\mathcal{F}(u), \mathcal{H}(u))=0,
\end{equation*}
for all $\mathcal{F}$. %
Such compatibility conditions imply
\begin{equation*}
  d\mathcal{H}(u) /dt = 0, \quad
  d\mathcal{S}(u) /dt = \big(\mathcal{S}, \mathcal{S}\big)(u) \leq 0,
\end{equation*}
that is, the entropy is dissipated at constant Hamiltonian. The qualitative idea is that, if $\mathcal{S}$ is bounded from below, the system will evolve on the manifold $H(u) = H(u_0)$, where $u_0$ is the initial condition, toward a state that satisfies $(\mathcal{S}, \mathcal{S}) = 0$. If the metric brackets vanish only in the direction of the Hamiltonian functional, i.e.,
\begin{equation}
  \label{eq:critical-condition}
  (\mathcal{S}, \mathcal{S}) = 0 
  \quad\Leftrightarrow\quad
  \frac{\delta \mathcal{S}}{\delta u} \propto \frac{\delta \mathcal{H}}{\delta u},
\end{equation}
then the relaxed state is a solution of the variational problem~(\ref{eq:energy_casimir_principle}). This is not always the case: some metric brackets have a larger ``null space'' so that the set of relaxed states strictly contains the solutions of~(\ref{eq:energy_casimir_principle}). If this happens we say that the operator does not completely \emph{control} the relaxation process.

\added[id=RW,remark=The above comment has been added in accordance with the referee question regarding the preservation of the topology of the initial condition]{In general the only constraint on the dynamics is that $u$ belongs to the manifold of constant energy $\mathcal{H}(u) = \mathcal{H}(u_0)$ The topology of the initial condition can be destroyed.}

\subsection{The metric operator and its applications to equilibria}

In this work we shall focus on the metric part of the dynamics and consider the class of metric brackets introduced by Morrison \cite{Morrison1984} as a generalisation of the Landau collision operator. Such operators will be referred to as integral collision-like operators. 

The general form for two arbitrary functionals $\mathcal{A}$ and $\mathcal{B}$  can be written as \cite{Morrison1984}
\begin{equation*}  
  (\mathcal{A}, \mathcal{B}) = - \iint  L_i \Bigl(\frac{\delta \mathcal{A}}{\delta u}\Bigr) 
  \cdot T_{i j} L_j \Bigl(\frac{\delta \mathcal{B}}{\delta u}\Bigr) dx dx^\prime,
\end{equation*} 
where $ L(v) = \nabla v(x) - \nabla v(x')$ and  $T_{i j} = T_{ij}(x,x')$ is a matrix with either scalar- or matrix-valued entries, depending on whether $u$ is a scalar or a multi-component field, respectively. Symmetry requires $T_{ij}(x,x^\prime) = T_{ji}(x^\prime, x)$. 

In order to ensure the conservation of a given Hamiltonian $\mathcal{H}$ we choose the kernel of the metric brackets according to
\begin{equation*}
T_{ij} (x, x^\prime) \propto 
    |g(x, x^\prime)|^2 I - g(x,x^\prime) \otimes g(x,x^\prime), 
\end{equation*} 
where $g = L ( \delta \mathcal{H} / \delta u )$.
A rigorous proof of~(\ref{eq:critical-condition}) for this class of operators is still not available. We shall however show in numerical experiments that the corresponding dynamics relaxes to a solution to~(\ref{eq:energy_casimir_principle}) as desired.

However this choice of the metric brackets leads to integral operators that are as challenging as the full Landau collision operator. Even though structure-preserving methods for the discretization of such operators are now available \cite{EeroAdams1, MichaelEero2}, we have introduced a simplified class of brackets leading to diffusion-like operators. Specifically we define
\begin{equation}\label{eq:final_form_diffusion_like_operators}
  (\mathcal{A}, \mathcal{B}) = - \int 
  \Big(\frac{\partial}{\partial x_i} \frac{\delta \mathcal{A}}{\delta u} \Big)
  \cdot D_{i j} \Big(\frac{\partial}{\partial x_i} \frac{\delta  \mathcal{B}}{\delta u} \Big)
  dx,
\end{equation}   
where $D(x) = |g(x)|^2 I - g(x) \otimes g(x)$ is an effective diffusion coefficient, with $g(x) = \nabla (\delta \mathcal{H} /\delta u)$.

\section{Computational Aspects}\label{sec:numerics}
For the time discretization the Crank-Nicolson scheme \cite{crank_nicolson_scheme} has been chosen, in order to guarantee discrete energy conservation, at least for quadratic energy functionals.
The discrete entropy is proven to be dissipated.
A finite element discretization has been chosen for the spatial operators. Continuous piecewise linear Lagrange elements have been used. %
Both the integro-differential operator and its local version have been implemented in FEniCS, a computing platform for solving partial differential equations \cite{fenics1, fenics2}.

\section{Numerical Experiments}\label{sec:numerical_experiments}
A gallery of different numerical experiments, for the models of Section \ref{sec:theory}, is presented here.

\subsection{2D Euler Equation}

The setup for the simulations is as follows: an anisotropic Gaussian is chosen as initial condition and Dirichlet homogeneous boundary conditions are applied for the squared domain $\Omega=[0,1]\times[0,1]$. \added[id=PS, remark=important specification of the resolution used]{A resolution of $64 \times 64$ points is chosen.} \added[id=RW,remark=The above comment has been added in accordance with the referee remark regarding the number of time steps]{A total number of $10000$ time steps, with a time step size of $100$, is simulated.} 

The entropy functional is quadratic in the dynamical variable $\omega$, namely \replaced[id=PS,remark=the choice of the entropy functional is stated more clearly]{$s(\omega) = \omega^2/2$ in equation \eqref{eq:euler_entropy_functional} }{$\mathcal{S} = ||\omega||^2 / 2$}%
, from which the variational principle predicts a linear relation between $\omega$ and the stream function $\phi$, $\omega = \lambda \phi$.  A local version of the collision-like metric operator, derived from equation \eqref{eq:final_form_diffusion_like_operators}, has been used to evolve the system. The energy functional is preserved with a relative precision of $10^{-12}$.

\begin{figure} \begin{center} %
\includegraphics[trim={0 0 0 0},clip,width=13cm,height=0.40\textwidth,keepaspectratio]{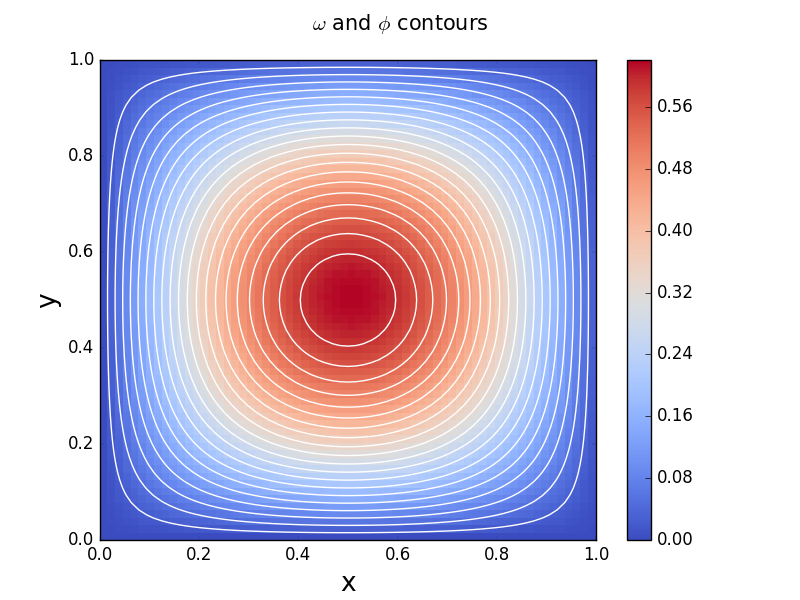}
\end{center}
\caption{\label{fig:euler_fields}\footnotesize{\added[id=PS,remark=testcase description added]{\textbf{Euler testcase, quadratic entropy}}: the color plot of $\omega$ together with the contours of $\phi$ at the equilibrium.}}
\end{figure}

Figure \ref{fig:euler_fields} shows the color plot of the vorticity field $\omega$ with the contours (represented with white solid curves) of the stream function $\phi$ at equilibrium, that is, at the end of the simulation. 
\replaced[id=PS,remark=better formulation]{At the equilibrium the contours of $\omega$ and $\phi$ should be the same and one can qualitatively see that this is the case in Figure \ref{fig:euler_fields}. A quantitative assessment of the equilibrium condition requires specific diagnostics.}{In order to verify whether the equilibrium condition selected by the choice of the entropy functional has been reached or not, it is useful to consider specific diagnostics.}
First of all, the time evolution of the entropy functional gives information about the relaxation process. When no appreciable variation of the entropy functional occurs, the system has reached an equilibrium. Moreover, it is interesting to make use of another diagnostic which \replaced[id=PS,remark=better explanation of the name chosen]{is a particular type of scatter plot}{in the following will be called a “scatter plot”}.
It is constructed by plotting for every grid node $(i, j)$ the corresponding discrete values $\phi_{i,j}$ and $\omega_{i,j}$: when the system is far from the equilibrium, 
\replaced[id=RW,remark=Clearer description as suggested by the referee]{these points are scattered over regions of the plane with no well-defined relation.}{these points are uniformly distributed on the cartesian plane.} %
On the other hand, as the system relaxes to the equilibrium, they show a functional relation, %
which can then be compared with what is theoretically expected from the variational principle. \deleted[id=PS,remark=redundant sentence]{The evolution of the entropy functional and the scatter plot of the system is a key diagnostics in order to understand whether an equilibrium condition has been reached or not.}

\begin{figure} \begin{center}  %
\includegraphics[trim={0 0 0 0},clip,width=13cm,height=0.40\textwidth,keepaspectratio]{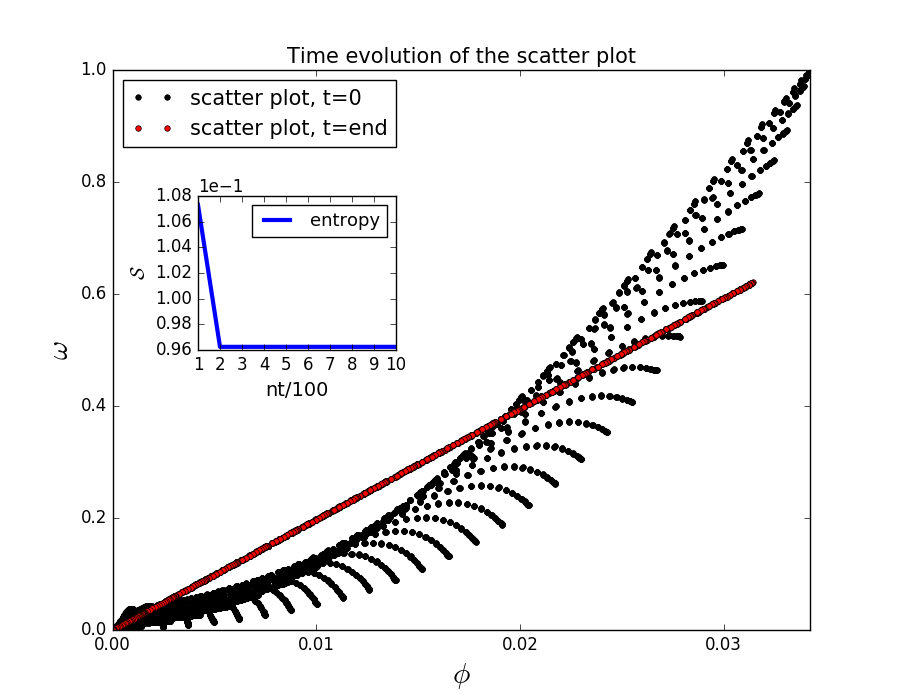}
\end{center}
\caption{\label{fig:euler_scatter}\footnotesize{\added[id=PS,remark=testcase description added]{\textbf{Euler testcase, quadratic entropy}}: a comparison of the functional relationship between $\omega$ and $\phi$ at the initial and final state. An inset shows the time evolution of the entropy functional. \added[id=RW,remark=The above comment has been added in accordance with the referee remark regarding the number of time steps]{The x-axis of the inset is in units of $100$ time steps.} As the entropy functional is minimised the system relaxes towards the equilibrium condition for which $\omega$ is a linear function of $\phi$.}}
\end{figure}

Figure \ref{fig:euler_scatter} shows the scatter plot at the beginning and at the end of the temporal evolution, together with the time evolution of the entropy functional. 

A fit of the functional relationship is performed, confirming that 
a linear functional relationship between $\omega$ and $\phi$ is found. \added[id=RW,remark=The above section has been added in accordance with the referee question regarding the error between the numerical and analytical eigenvalue]{Two verification tests of the numerical results have been then carried out. 
In the case of the choice of a quadratic entropy functional analytical results are available. }

In fact equation \eqref{eq:euler_equilibrium}, which describes the physical equilibria, reduces to the linear eigenvalue problem
\begin{equation} \label{eq:linear_eigenvalue_problem}
-\Delta \phi = \lambda \phi,
\end{equation}
which can be solved analytically. The analytical eigenvalues can be compared with the result of the fit in the numerical simulation.

\begin{figure} \begin{center}  %
\includegraphics[trim={0 0 0 0},clip,width=8.5cm,keepaspectratio]{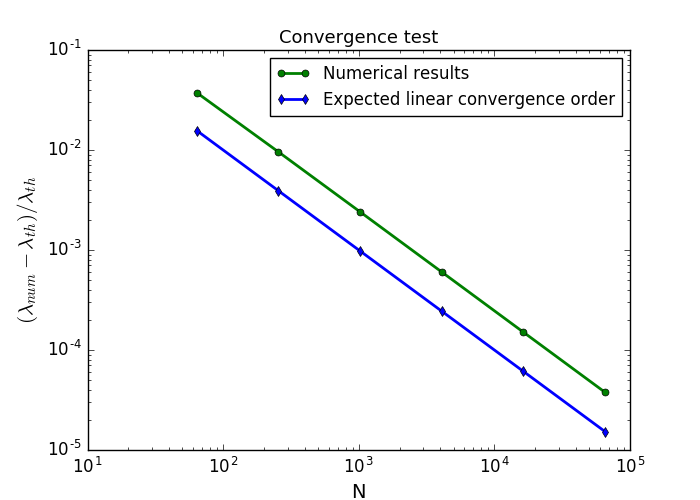} 
\end{center}
\caption{\label{fig:convergence_test}\footnotesize{\textbf{Euler testcase, quadratic entropy}: convergence test in a logarithmic scaling: the relative error between the numerical and analytical eigenvalue is plotted against different resolutions. The convergence order is tested against a reference line of order one.}}
\end{figure}

The eigenvaues in a domain $\Omega=[0,a]\times[0,b]$ with Dirichlet boundary conditions are
\begin{equation} \label{eq:laplace_operator_eigenvalues}
\lambda_{n , m} = \pi^2 \Bigl(\Bigl(\frac{n}{a}\Bigr)^2 + \Bigl(\frac{m}{b} \Bigr)^2 \Bigr), \quad n,m \geq 1 .
\end{equation}
In our example $a=b=1$, and thus the analytical eigenvalue corresponding to the fundamental state (the equilibrium) is $\lambda_{1,1} = 2 \pi^2 \approx 19.73921$. %
\added[id=RW,remark=The above section has been added in accordance with the referee question regarding the error between the numerical and analytical eigenvalue]{
Figure \ref{fig:convergence_test} shows a convergence test in which the relative error between the numerical and the analytical eigenvalue is plotted against increasing mesh resolutions in a logarithmic scaling. A numerical fit shows a convergence order equal to $1$. This result is in agreement with the choice of the order of the finite elements used, namely piecewise linear Lagrange elements.}

\added[id=RW,remark=The above section has been added in accordance with the referee question regarding the error between the numerical and analytical eigenvalue]{The second verification test has been performed with another numerical algorithm, described in \cite{TakedaTokuda}, to compute the fundamental eigenvalue of equation \eqref{eq:linear_eigenvalue_problem}. %
The maximum relative error between the result of the numerical fit and the value computed with this procedure for each mesh resolution is of the order $10^{-8}$, using the same spatial discretization method. %
} %

Other numerical experiments confirm that the equilibrium reached, given a choice for the boundary conditions, is independent of the initialisation chosen for the simulation (initial conditions), being driven by the choice of the entropy functional only. \added[id=RW,remark=The above comment has been added in accordance with the referee question regarding the preservation of the topology of the initial condition]{A topology change of the initial configuration can also occur.} All the numerical results \replaced[id=PS,remark=more correct sentence as not for every testcase a convergence study of the error has been carried out]{behave in the same way as the case for which a convergence test has been performed}{are found in agreement with the analytical solution}.    \\ %

The same test case has been simulated in a complicated mesh, with the domain constructed from a unitary circle mapped by the Czarny mapping \cite{czarny}. \added[id=PS, remark=important specification of the resolution used]{A resolution of $8270$ points is chosen. %
} \added[id=RW,remark=The above comment has been added in accordance with the referee remark regarding the number of time steps]{A total number of $220000$ time steps, with a time step size of $100$, is simulated.} %

\begin{figure} \begin{center} %
\includegraphics[trim={0 0 0 0},clip,width=13cm,height=0.40\textwidth,keepaspectratio]{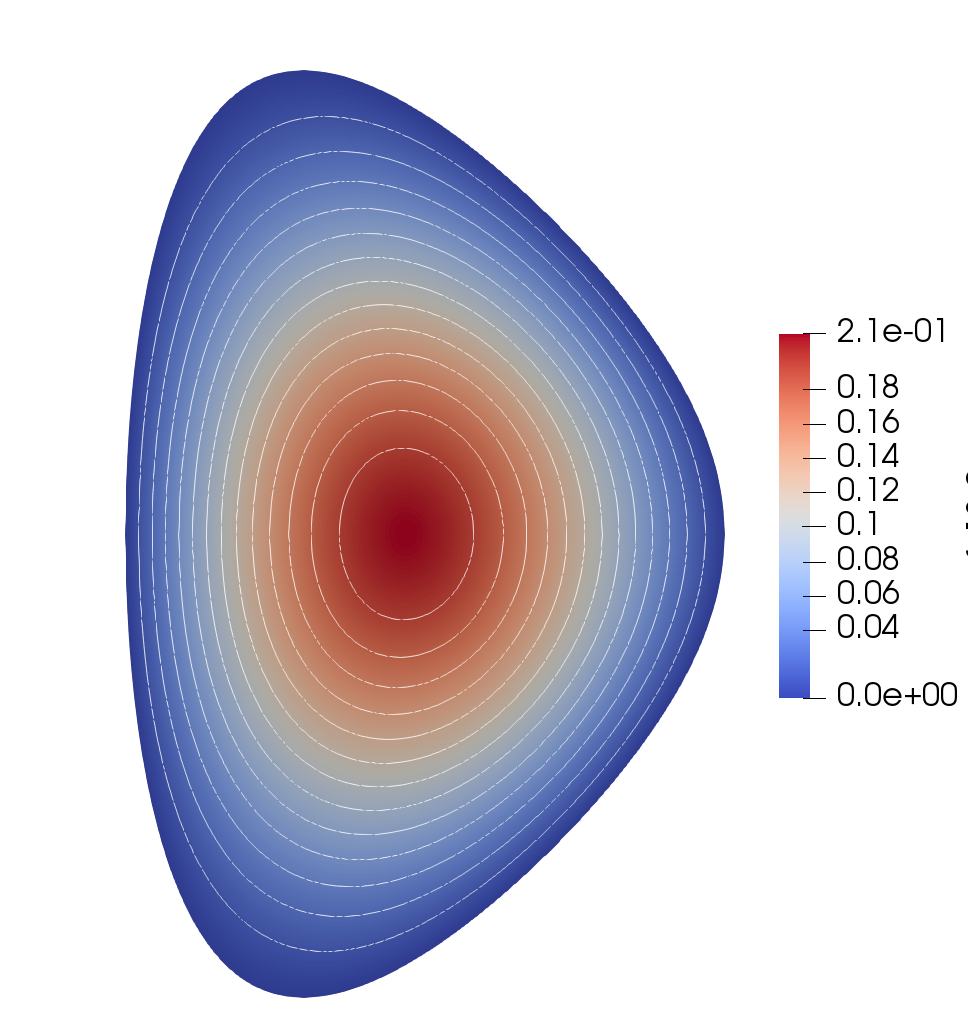}
\end{center}
\caption{\label{fig:euler_czarny_fields}\footnotesize{\added[id=PS,remark=testcase description added]{\textbf{Euler testcase, quadratic entropy in a Czarny mapped domain}}: the color plot of $\omega$ with the contours of $\phi$ at the equilibrium.}} \end{figure}

Figure \ref{fig:euler_czarny_fields} shows the color plot of the dynamical variable $\omega$ and the solid white lines representing the contours of the streaming function $\phi$ at equilibrium.

\begin{figure} \begin{center} %
\includegraphics[trim={0 0 0 0},clip,width=13cm,height=0.40\textwidth,keepaspectratio]{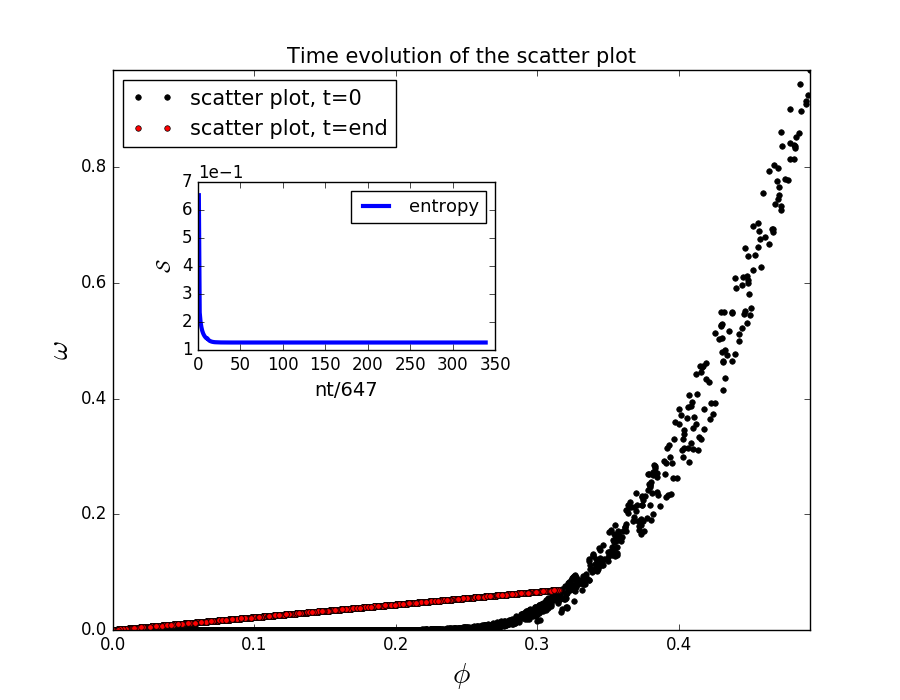}
\end{center}
\caption{\label{fig:euler_czarny_scatter}\footnotesize{\added[id=PS,remark=testcase description added]{\textbf{Euler testcase, quadratic entropy in a Czarny mapped domain}}: same as in Figure \ref{fig:euler_scatter}. \added[id=RW,remark=The above comment has been added in accordance with the referee remark regarding the number of time steps]{The x-axis of the inset is in units of $647$ time steps.} The mapped domain does not play any role in the selection of the physical equilibrium.}} 
\end{figure}

In Figure \ref{fig:euler_czarny_scatter} the scatter plot at the initial and equilibrium state of the simulation and the time evolution of the entropy functional are shown. Again as the system reaches the equilibrium state, i.e. as the entropy functional is minimised, the expected linear functional relationship between $\omega$ and $\phi$ appears. \added[id=PS, remark=numerical eigenvalue added for consistency]{The numerical eigenvalue can be computed from the final functional relationship with a fit. The result is $\lambda_{\text{num}} \approx  0.2203$, with a fit error of  $10^{-9}$. %
Due to the non trivial domain, it is not possible to verify this result against an analytical value. Therefore, the only verification test available is with the numerical algorithm described in \cite{TakedaTokuda}: the relative error between the two numerical results is $10^{-6}$.}

The more complicated domain does not influence the selection of the physical equilibrium, which is due to the choice of the entropy functional only. The method can thus be applied in complicated domains.

\subsection{Grad Shafranov}
Numerical simulations for the Grad Shafranov model have also been performed. The simulation setup used here is the following: anisotropic Gaussian as initial condition, homogeneous Dirichlet boundary conditions on a rectangular domain $\Omega=[1.0,7.0]\times[-9.5,9.5]$. \added[id=PS, remark=important specification of the resolution used]{A resolution of $128 \times 128$ points is chosen.} %
 \added[id=RW,remark=The above comment has been added in accordance with the referee remark regarding the number of time steps]{A total number of $53000$ time steps, with a time step size of $100$, is simulated.}
The entropy functional is chosen according to the equilibrium to be selected. %
As in the simulations for the Euler case, the energy functional is preserved \replaced[id=PS,remark=corrected syntax]{with a relative error of $10^{-13}$ or smaller.}{with up to a relative precision of $10^{-13}$}

As a first example, the entropy functional in equation \eqref{eq:gs-hm_entropy_functional} has been chosen in order to reproduce the Herrnegger-Maschke solution described in Section \ref{sec:theory}. The functional relation between  $u=(4\pi/c) \: R j_{\phi}$ and $\psi$ is $u = \lambda (C R^2 + D) \psi$, \added[id=RW,remark=The above comment has been added in accordance with the referee question regarding the numerical values of $C$ and $D$]{where $C$ and $D$ are arbitrary positive constants equal to $0.6$ and $0.18$ respectively.} %

\begin{figure} \begin{center} %
\includegraphics[trim={0 0 0 0},clip,width=13cm,height=0.40\textwidth,keepaspectratio]{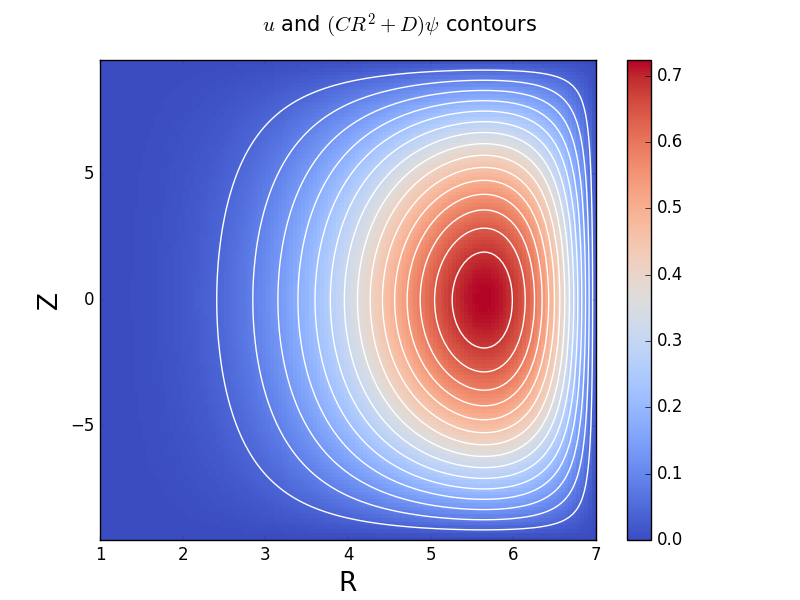}
\end{center}
\caption{\label{fig:gs_HMentropy_fields}\footnotesize{\added[id=PS,remark=testcase description added]{\textbf{Grad-Shafranov testcase, Herrnegger-Maschke entropy:}} the color plot of $u$ and the contours of $(C R^2 + D) \psi$ at equilibrium.}} \end{figure}

In Figure \ref{fig:gs_HMentropy_fields} the color plot of the dynamical variable is shown together with the contours of the flux function $\psi$. The contours of the two fields are aligned, as the simulation has reached the equilibrium state. 

\begin{figure} \begin{center} %
\includegraphics[trim={0 0 0 0},clip,width=13cm,height=0.40\textwidth,keepaspectratio]{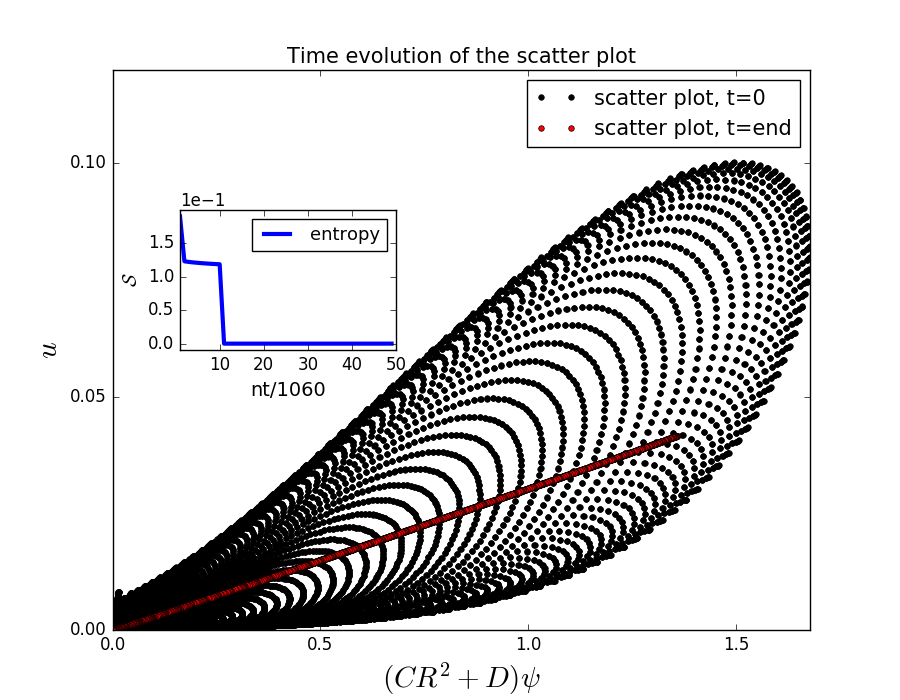}
\end{center}
\caption{\label{fig:gs_HMentropy_scatter}\footnotesize{\added[id=PS,remark=testcase description added]{\textbf{Grad-Shafranov testcase, Herrnegger-Maschke entropy:}} the functional relationship between the dynamical variable $u$ and $(C R^2 + D) \psi$ is shown at the initial and final state, together with the temporal evolution of the entropy functional. \added[id=RW,remark=The above comment has been added in accordance with the referee remark regarding the number of time steps]{The x-axis of the inset is in units of $1060$ time steps.} At convergence, $u$ is a linear function of $(C R^2 + D) \psi$, as expected.}} \end{figure}

A more quantitative analysis of the equilibrium state is shown in Figure \ref{fig:gs_HMentropy_scatter}. Here the functional relationship between the dynamical variable \replaced[id=RW,remark=Correct relationship as pointed out by the referee]{$u$ and $(CR^2 + D)\psi$}{$u$ and $\psi$} is shown at two different simulation times: at the initial and final state. An inset shows the time evolution of the entropy functional, which is minimised. The functional relationship at convergence is the one expected from the variational principle, and the numerical eigenvalue can be computed with a fit. \added[id=PS, remark=added numerical eigenvalue and comparison with the result of the recursive algorithm of Takeda and Tokuda]{The result of the fit is $\lambda_{\text{num}} \approx 0.0305$, with a fit error of $10^{-6}$. %
A verification test against the numerical result given by the algorithm described in \cite{TakedaTokuda} shows agreement with a relative error of $10^{-3}$.  }

An entropy functional quadratic in the dynamical variable $u$, \replaced[id=PS,remark=the choice of the entropy functional is stated more clearly]{$s(u, R, z) = u^2 / 2$ in equation \eqref{eq:gs_entropy_functional}  }{$||u||^2 / 2$}, has been chosen in order to run the same simulation on a Czarny mapped domain \cite{czarny}.  \added[id=PS, remark=important specification of the resolution used]{A resolution of $8270$ points is chosen. %
}  \added[id=RW,remark=The above comment has been added in accordance with the referee remark regarding the number of time steps]{A total number of $18000$ time steps, with a time step size of $100$, is simulated.} The initial configuration is otherwise the same as stated in the previous testcase.
The variational principle still describes a linear functional relation between $u$ and the flux function $\psi$, $u = \lambda \psi$.

\begin{figure} \begin{center} %
\includegraphics[trim={0 0 0 0},clip,width=13cm,height=0.40\textwidth,keepaspectratio]{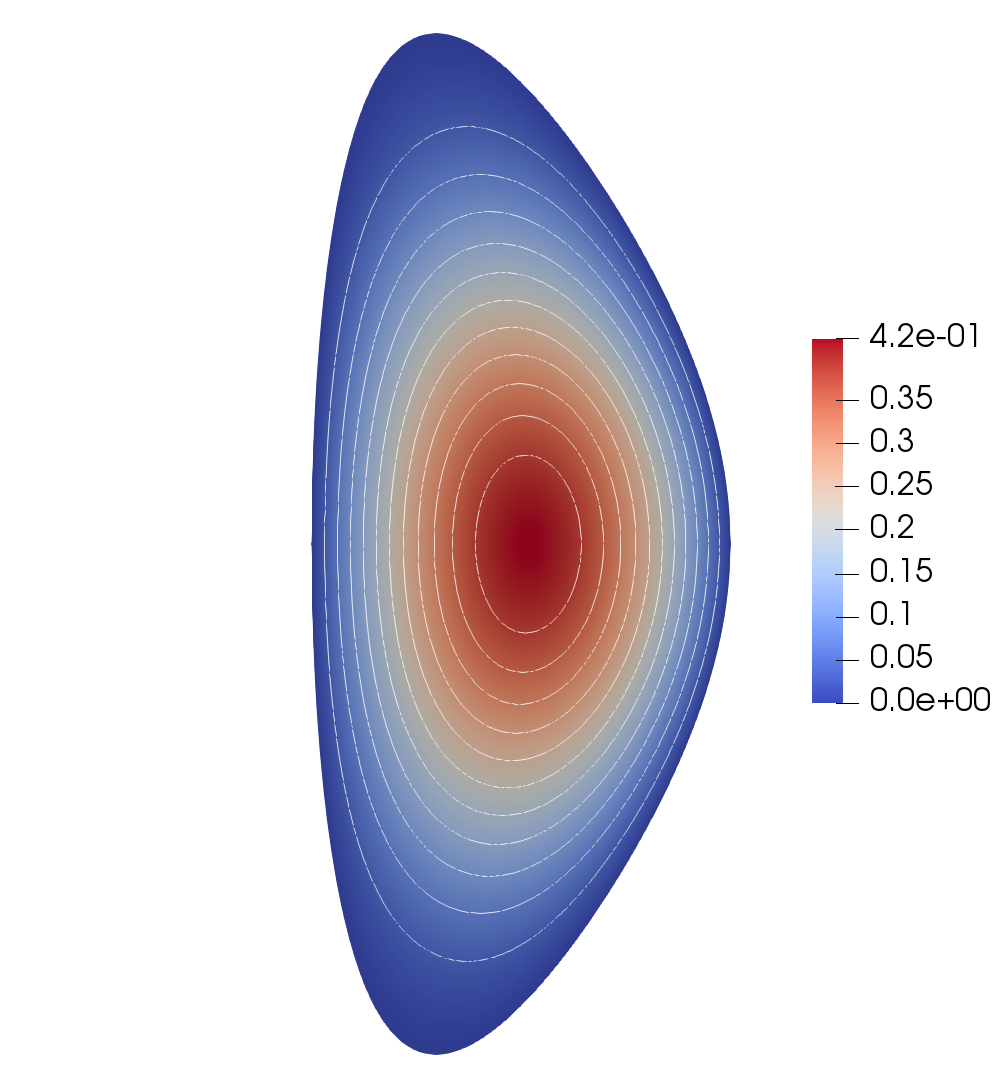}
\end{center}
\caption{\label{fig:gs_czarny_fields}\footnotesize{\added[id=PS,remark=testcase description added]{\textbf{Grad-Shafranov testcase, quadratic entropy in a Czarny mapped domain:}} the color plot of $u$ and the contours of $\psi$ at equilibrium.}} \end{figure}

Figure \ref{fig:gs_czarny_fields} shows the color plot of $u$ with the white solid lines representing the contours of $\psi$.

\begin{figure} \begin{center} %
\includegraphics[trim={0 0 0 0},clip,width=13cm,height=0.40\textwidth,keepaspectratio]{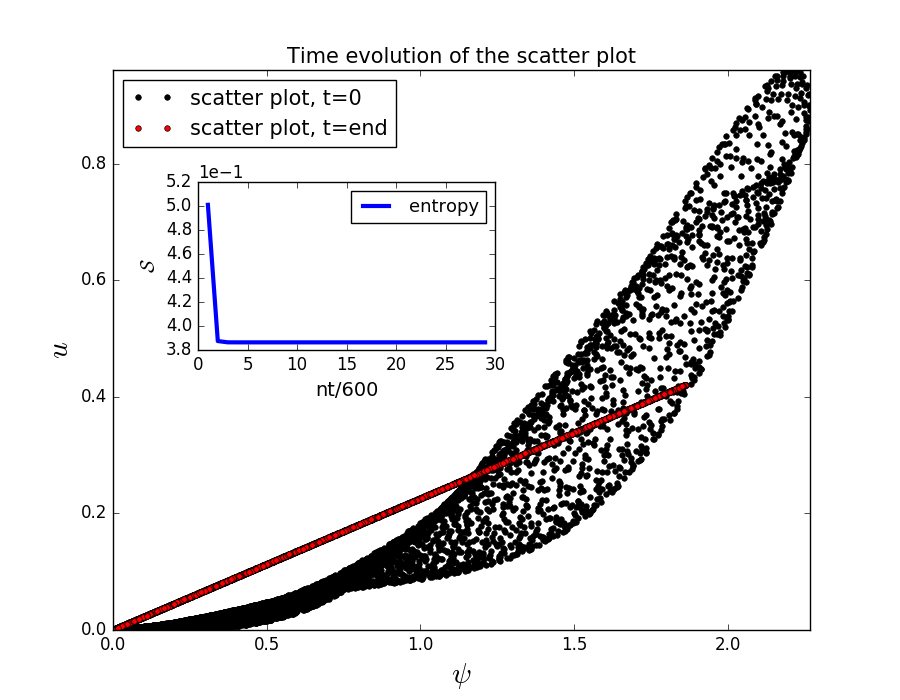}
\end{center}
\caption{\label{fig:gs_czarny_scatter}\footnotesize{\added[id=PS,remark=testcase description added]{\textbf{Grad-Shafranov testcase, quadratic entropy in a Czarny mapped domain:}} same as in Figure \ref{fig:gs_HMentropy_scatter}. \added[id=RW,remark=The above comment has been added in accordance with the referee remark regarding the number of time steps]{The x-axis of the inset is in units of $600$ time steps.}}} \end{figure}

Figure \ref{fig:gs_czarny_scatter} shows the scatter plot diagnostics at the initial and final state. As in the previous cases, an inset shows the time evolution of the entropy functional.  
As the entropy functional is minimised, the functional relation collapses to a linear function of the two variables $u$ and $\psi$, as expected from the variational principle. \added[id=PS, remark=numerical eigenvalue added for consistency]{The numerical eigenvalue is again computed: the result is  $\lambda_{\text{num}} \approx  0.226$ with a fit error of $10^{-8}$. %
The relative error between this result and the one obtained by the algorithm described in \cite{TakedaTokuda} is $10^{-7}$. }  \newline

\section{Outlook}
\added[id=RW,remark=The above section has been added in accordance with a referee recommendation]{
In this work we have shown, in simple testcases, that metriplectic dynamics can be used as a  relaxation method for the calculation of equilibria.
The method requires a significant number of iterations in order to relax to the equilibrium and for the two-dimensional models considered here, it is not competitive as compared to standard approaches. However it does have the advantage of being applicable to generic three-dimensional equilibria, as long as a variational principle of the form  \eqref{eq:energy_casimir_principle} is available. This is the case for force-free MHD equilibria (Beltrami fields discussed in Section \ref{subsection:variational_principle}).
The application of the method to full three-dimensional ideal MHD equilibria is currently under investigation. The main difficulty consists in recasting the equilibrium problem in the form of equation \eqref{eq:energy_casimir_principle} with appropriate energy and entropy functionals. The standard variational formulation for MHD equilibria \cite{chodura_schlueter} cannot be directly exploited since it makes use of constrained variations. %
The appropriate reformulation of the variational principle has to be addressed. } \\

%
%

\textit{ We acknowledge useful discussions with Yaman ~G\"u\c{c}l\"u and Edoardo ~Zoni regarding the Euler equilibrium domain mapping. } %

\textit{ This work has been carried out within the framework of the EUROfusion Consortium and has received funding from the Euratom research and training programme 2014-2018 under grant agreement No 633053. The views and opinions expressed herein do not necessarily reflect those of the European Commission. } %

\added[id=PS, remark=coauthor support statement added]{\textit{P. J. M. was supported by the US Dept.of Energy Contract $\#$DE-FG05-80ET-53088, a Forschungspreis from the Humboldt Foundation, and in part by the National Science Foundation under Grant No. DNS-1440140 while he was in residence at the Mathematical Science Research Institute in Berkeley, CA during the Fall 2018 semester.}}

\listofchanges[style=summary]


\section*{References}
\begin{thebibliography}{9}
\bibitem{vmec1} Hirschman S P and Whitson J C 1983, {\it Phys. Fluids}, {\bf 26}, 3553
\bibitem{vmec2} Hirschman S P and Whitson J C 1986, {\it Comput. Phys. Commun.}, {\bf 43}, 143
\bibitem{pies} Reiman A H and Greenside H S 1986, {\it Comput. Phys. Commun.}, {\bf 43}, 157
\bibitem{siesta} Hirshman \etal. 2011, {\it Phys. Plasmas}, {\bf 18}, 062504
\bibitem{spec} Hudson S R, Dewar R L \etal. 2012, {\it Plasma Phys. Controll. Fusion}, {\bf 54}, 014005 
\bibitem{Morrison1984} Morrison P J 1984, {\it Phys. Lett. A}, {\bf 100}, 423-7
\bibitem{Morrison1986} Morrison P J 1986, {\it Physica D}, {\bf 18}, 410-9
\bibitem{MaterassiMorrison} Materassi M and Morrison P J 2018, {\it J. Cybernetics and Physics} , Accepted 
%
%
\bibitem{MaterassiAndTassi} Materassi M and Tassi E, {\it Physica D}, {\bf 241}, 6
\bibitem{MittnenzweigAndMielke} Mittnenzweig M and Mielke A 2017, {\it J. of Statistical Physics}, {\bf 167}, 205-233
\bibitem{PhilReviewPaper} Morrison P J 1998, {\it Rev. Mod. Phys.}, {\bf 70}, 2
\bibitem{Gay-BalmazHolm} Gay-Balmaz F and Holm D 2013, {\it Nonlinearity}, {\bf 26}, 495-524 %
\bibitem{Guy-Balmaz_selectivCasimir_MHD} Gay-Balmaz F and Holm D 2014, {\it Nonlinearity}, {\bf 27}, 1747-1777
\bibitem{FlierlAndMorrison} Flierl G R and Morrison P J 2011, {\it Physica D}, {\bf 240}
\bibitem{ChikasueFurukawa1} Chikasue Y and Furukawa M 2015 {\it Phys. Plasmas}, {\bf 22}, 02251
\bibitem{ChikasueFurukawa2} Chikasue Y and Furukawa M 2015 {\it J. Fluid Mec.}, {\bf 744}, 443-59
\bibitem{FurukawaMorrison3} Furukawa M and Morrison P J 2017 {\it Plasma Phys. Controll. Fusion}, {\bf 59}, 054001 %
\bibitem{FurukawaMorrison4} Furukawa M, Morrison P J, Watanabe T and Ichiguchi T 2018, {\it Phys. Plasmas}, {\bf 25} 082506 
\bibitem{Lenard} Lenard A 1900, {\it Ann. Phys.}, {\bf 3}, 390-400
%
%
\bibitem{GS1} Shafranov V D 1958, {\it Sov. Phys. JETP}, {\bf 6}, 545
\bibitem{GS2} Grad H and Rubin H 1958, {\it Proc. of the 2nd United Nations
Conf. on the Peaceful uses of Atomic Energy}  ~United Nations, Geneva vol. 21, p.190
\bibitem{GS_standard_VP} Lao L L, Hirschman S P, Wieland R M 1981, {\it American Institute of Physics}, {\bf 24}, 1431
\bibitem{GSMcCarthypaper} Mc Carthy P J 1999, {\it Phys. Plasmas}, {\bf 6}, 3554
\bibitem{beltrami1} Woltjer L 1958, {\it Proc. of the National Academy of Sciences}, {\bf 44}, 6
\bibitem{beltrami2} Taylor J B 1974, {\it Phys. Rev. Lett.}, {\bf 33}, 19, 1139-1141
%
\bibitem{leibniz_jacobi} Bloch A M, Morrison P J and ratiu T S 2013, {\it Springer Proc. in Mathematics \& Statistics}, {\bf 35}, pp. 371–415
%
\bibitem{EeroAdams1} Hirvijoki E and Adams M F 2017, {\it Phys. Plasmas}, {\bf 24}, 032121
\bibitem{MichaelEero2} Kraus M and Hirvijoki E 2017, {\it Phys.  Plasmas}, {\bf 24}, 102311
\bibitem{crank_nicolson_scheme} Crank J, Nicolson P 1947, {\it Proc. Camb. Phil. Soc.}, 43 (1), 50-67
\bibitem{fenics1} Martin Aln{\ae}s S \etal. 2015.  %
{\it Archive of Numerical Software}{ \bf 3}, 100, 9-23
\bibitem{fenics2} Logg A, Mardal K A, Wells G N \etal. 2012, {\it Automated Solution of Differential Equations by the Finite Element Method}, (Springer) %
%
%
\bibitem{TakedaTokuda} Takeda T and Tokuda S 1991, {\it J. Comp. Phys.}, {\bf 93}, 1-107
\bibitem{czarny} Czarny O and Huysman G 2008, {\it J. Comp. Phys.}, {\bf 16}, 227, 7423-7445
%
\bibitem{chodura_schlueter} Chodura R and Schl\"uter A 1981, {\it J. Comp. Phys. }, {\bf 41}, 68-88
%



%

%
\end{thebibliography}
\end{document}